# Wide-Range MRI Artifact Removal with Transformers

Lennart Alexander Van der Goten[1,2]
lavdg@kth.se

Kevin Smith[1,2]
ksmith@kth.se

[1] KTH Royal Institute of Technology
Stockholm, SWEDEN

[2] Science for Life Laboratory
Solna, SWEDEN

*for the Alzheimer's Disease Neuroimaging Initiative*[*]

### Abstract

Artifacts on magnetic resonance scans are a serious challenge for both radiologists and computer-aided diagnosis systems. Most commonly, artifacts are caused by motion of the patients, but can also arise from device-specific abnormalities such as noise patterns. Irrespective of the source, artifacts can not only render a scan useless, but can potentially induce misdiagnoses if left unnoticed. For instance, an artifact may masquerade as a tumor or other abnormality. *Retrospective artifact correction* (RAC) is concerned with removing artifacts after the scan has already been taken. In this work, we propose a method capable of retrospectively removing eight common artifacts found in native-resolution MR imagery. Knowledge of the presence or location of a specific artifact is not assumed and the system is, by design, capable of undoing interactions of multiple artifacts. Our method is realized through the design of a novel volumetric transformer-based neural network that generalizes a *window-centered* approach popularized by the Swin transformer. Unlike Swin, our method is (i) natively volumetric, (ii) geared towards *dense prediction* tasks instead of *classification*, and (iii), uses a novel and more global mechanism to enable information exchange between windows. Our experiments show that our reconstructions are considerably better than those attained by ResNet, V-Net, MobileNet-v2, DenseNet, CycleGAN and BicycleGAN. Moreover, we show that the reconstructed images from our model improves the accuracy of FSL BET, a standard skull-stripping method typically applied in diagnostic workflows.

## 1 Introduction

Artifacts are a topic of tremendous practical significance within radiology. Patient movement within an MRI tube can result in so-called *motion artifacts* that severely degrade the quality of a scan, often impairing further downstream analyses or prompting a costly re-scan. Other artifacts can arise due to the nature of the MRI device, *e.g.* due to inhomogeneities of the magnetic field, noise or perturbations in the frequency domain. *Artifact correction* (AC) is the process that is concerned with removing artifacts from imagery while retaining medically-relevant information. Approaches to artifact removal can be grouped into two subdisciplines: *Prospective* artifact correction methods (PAC) require the collection of additional information (*e.g.* motion tracking of the patient) to undo artifacts, while *retrospective* artifact correction methods (RAC) operate solely on the scan produced by the device without any additional data gathering.

[*]Data used in preparation of this article were obtained from the Alzheimer's Disease Neuroimaging Initiative (ADNI) database (adni.loni.usc.edu). As such, the investigators within the ADNI contributed to the design and implementation of ADNI and/or provided data but did not participate in analysis or writing of this report. A complete listing of ADNI investigators can be found at: http://adni.loni.usc.edu/wp-content/uploads/how_to_apply/ADNI_Acknowledgement_List.pdf

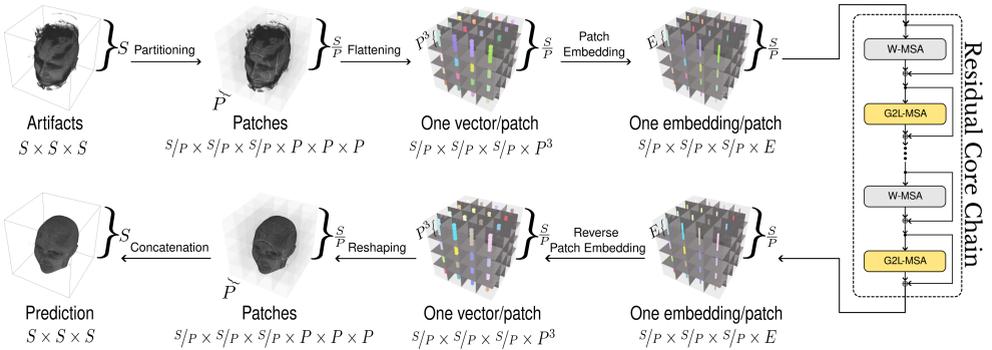

Figure 1: *Overview W-G2L-ART:* We describe a mirrored approach to tackle retrospective artifact correction in volumetric image data. An 3D scan exhibiting artifacts is first dissected into patches. Each flattened patch gets embedded into an $E$-dimensional space using a shared fully-connected layer. The heavy-lifting is done by the *Residual Core Chain*, a chain of residual blocks that *alternate* between two window attention types, W-MSA (adapted from Swin) and G2L-MSA (proposed here), with the latter being specifically designed to allow fully-global window-to-window communication. The reverse path reshapes the data to produce dense predictions of a cleaned volumetric scan.

In this work, we propose the Windowed Global-to-Local Artifact Removal Transformer (*W-G2L-ART*, see Figure 1), that operates retrospectively on volumetric data (*i.e.* not on individual slices [21]) and is capable of reverting a rich class of artifacts without requiring information as to which specific artifacts are present. We simulate artifacts, like prior works [7, 8, 20, 51, 55, 53], but instead of assuming that only a single artifact is present we use a *Bernoulli process* during training to sample artifacts independently, potentially resulting in multiple artifacts per scan. This enables us to streamline both the model architecture and the training process, as we can cast the learning task as a supervised problem for which paired data is readily available.

The recent success of vision transformers [6] has popularized neural networks that no longer have convolutional layers as their main building blocks, but instead rely on attention mechanisms. Motivated by the success of Swin [21], we investigate here whether a window-based transformer architecture is beneficial in the context of RAC. To this end we devise an approach centered around volumetric windows, but propose to replace the *shifted-window* attention (SW-MSA) of Swin with a new mechanism, which we call *global-to-local* attention (G2L-MSA). In a nutshell, we make the following contributions:

1. We propose W-G2L-ART, an artifact correction system that enables the reversal of a diverse set of artifacts often encountered in practice

2. We suggest a window-based self-attention mechanism that, like Swin (SW-MSA), uses

local windows to make vision transformers efficient. But our mechanism, G2L-MSA, is more general and flexible, allowing long-range connections in each window if desired. It can be used as a drop-in replacement for SW-MSA and is easily extended to 3D

3. Our experiments show that W-G2L-ART outperforms volumetric variants of the ResNet, V-Net, MobileNet-v2, DenseNet, CycleGAN and BicycleGAN. Our model also compares favorably against W-SW-MSA-ART, an ablation of our model that uses a 3D variant of Swin attention (SW-MSA).

4. We demonstrate that scans reconstructed by our model lead to improvements w.r.t. a skull stripping task performed with FSL BET.

## 2   Related work

Artifact removal can be categorized as a type of image restoration which, in turn, is a subfield of the *inverse* problem class. Image restoration is mainly concerned with deblurring, denoising and super-resolution of photographic images, but has applications in medicine and microscopy as well. Instead of handcrafted filters, the deep learning era has given rise to *learnable* models that no longer need *a priori* knowledge of the underlying degradation. Early examples include DeblurGAN and its successor for deblurring [15, 16], super-resolution networks [17] as well as denoising [36].

The image restoration application of concern in this work is retrospective artifact correction on MRI scans. Earlier works in this area concentrated on motion artifacts, including Haskell *et al.* [7], who propose TAMER to estimate motion as a rigid process and solve the corresponding non-linear equations to filter out the artifacts. A follow-up architecture called NAMER [8] follows a similar approach but relies on convolutional layers to perform the estimation. SARA-GAN [35] is a generative adversarial network for compressive sensing on undersampled MR images to reduce scanning times while retaining reasonable quality. Usman *et al.* [31] also experiment with adversarial learning, but in a *multi-shot*\* scan setting.

More recent work (e.g. [13, 27]) tackles the RAC problem within the fMRI resp. ultrasound domain but are not easily portable to the MRI domain.

Artifact types other than motion have received less attention from the community. Tustison *et al.* [29] propose to remove bias fields by nonparametric intensity normalization, whereas Gaillochet *et al.* [6] couple bias field correction with undersampled *k*-space reconstruction within an unsupervised context. Ghosting artifacts, which are also caused by patient movement, can be reduced via iterative inverse problem solving [28] if motion parameters are known or retrospectively by using ALOHA [13].

A recent work, DUNCAN [20], formulates RAC as an adversarial game between two generator-discriminator pairs that aim to translate between an artifact-free and artifact-defining domain by enforcing *cyclic consistency* among the two generators. In this work, corrections are restricted to those caused by motion, namely, reflecting background noise movement, swallowing, and other sudden movements [4]. In all the prior works including [20] artifacts are simulated.

Our approach synthesizes a wide array of artifacts that better reflects real-world data, including (but not limited to) bias field inhomogeneities, ghosting, and herringbone artifacts.

---

\*Refers to the process of acquiring a (single) *k*-space using scans performed at different time points

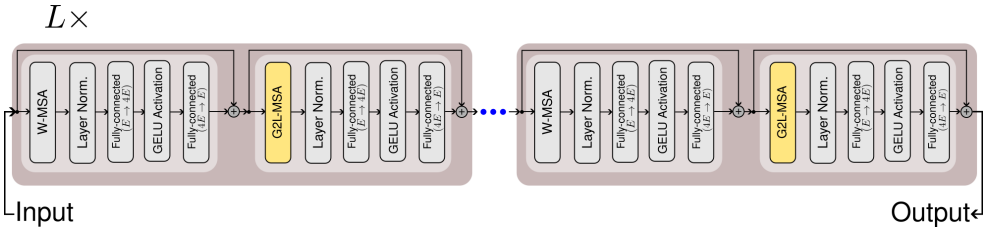

Figure 2: *Residual Core Chain*: The design of our suggested artifact-removal model, named W-G2L-ART, relies on the definition of a chain of *L* blocks (c.f. Figure 1). Each block, in turn, contains two sub-blocks that only differ in the attention mechanism they utilize. The first attention mechanism is the (local) W-MSA initially popularized by the Swin transformer [21]. We suggest a second attention mechanism, called G2L-MSA, to perform a windowed attention computation that factors in a more extensive *global* context than the corresponding SW-MSA operation used by Swin.

Furthermore, instead of operating slice-by-slice, our model applies a full *volumetric* prediction to increase the expressiveness of the model and enforce smooth intra-slice transitions. Thirdly, we employ a *Bernoulli* sampling process to corrupt images with a random composition of artifacts, providing a wider support coverage. Finally, we opt for a more streamlined supervised learning approach using 3D transformers instead of CNNs, which we motivate by our belief that the global nature of G2L-MSA could benefit the reversal of artifacts that are spatially spread out.

Our work takes inspiration from the Swin transformer [21] which uses windowed attention mechanisms to make classification of high-resolution (2D) imagery tractable. Recently, SwinIR [19] generalized Swin to dense predictions on natural images (2D) for super-resolution, denoising and JPEG artifact reduction. Importantly, SwinIR does not fully abstain from convolutional elements and is trained on a per-task basis, furthermore, it relies on the same attention mechanisms as Swin. To the best of our knowledge, there is no previous work that addresses a wide range of artifacts in 3D data, nor one that applies vision transformers to volumetric image reconstruction.

## 3 Method

Our task is to filter out a wide range of artifact types from volumetric image scans. To accomplish this, we propose W-G2L-ART (see the overview in Figure 1). Simulated artifacts are added to high-fidelity scans, and a dense prediction 3D vision transformer network with novel *Residual Core* blocks is trained to remove the artifacts.

Given a set of 3D MR scans $(X^{(i)})_{i=1,...,N} \in \mathbb{R}^{S \times S \times S}$ where $S$ denotes the resolution (*e.g.* $S = 256$ for a voxel size of 1mm$^3$), we are interested in *filtering out* artifacts from a wide range of artifact types $\mathcal{A}_1, ..., \mathcal{A}_K$. As artifacts often represent irreversible degradations, we assume that there is a not necessarily unique function $f$ that maps a scan $X$ onto its artifact-free representation $\tilde{X} \in \mathbb{R}^{S \times S \times S}$. Throughout this work, we assume that data used for training is supplied as pairs $(X, \tilde{X})$ where $X$ is derived from the original $\tilde{X}$ by leveraging an *artifact simulator* that stochastically samples a series of artifacts from $\mathcal{A}_1, ..., \mathcal{A}_K$ that are applied *sequentially* onto $\tilde{X}$ using a *Bernoulli* process.

**Considerations** Vision transformers [32] have recently become a staple of computer vision.

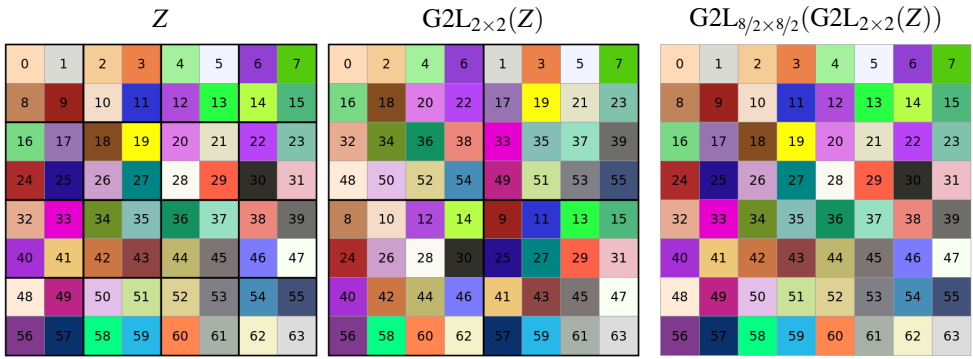

Figure 3: *Global-to-Local (G2L) Function*: *(left)* An input matrix $Z \in \mathbb{R}^{S_1 \times S_2}$ (here $S = S_1 = S_2 = 8$) is decomposed by the suggested $G2L_{U \times V}$ function into a grid of non-overlapping regions (framed by **black** borders) of size $U \times V$ (here $U = V = 2$). *(center)* The result is then defined by collecting values with identical local region coordinates (*e.g.* the elements $0, 2, 4, 6, 16, 18, \ldots, 54$ have the same local region coordinate $(0, 0)$ in *Z*). *(right)* The inverse function of $G2L_{U \times V}$ is given by $G2L_{S_1/U \times S_2/V}$ (here $G2L_{4 \times 4}$) and restores *Z* upon application. Values are color-mapped to improve readability. The G2L function class can best be envisioned as an injective mapping between spatial positions.

In this work, we adapt ideas from 2D natural image vision transformers to enable us to generate volumetric dense predictions while still reaping the benefits of the conventional transformer architecture. In a similar fashion to ViT [3], we partition the image (or 3D *scan*) into *patches* that are volumetric in our case. One challenge of volumetric predictions in terms of tractability is that the number of voxels increases *cubically* with $S$. Naively generalizing ViT to 3D would incur a sequence length of $(S/P)^3$ in the *self-attention* module, as opposed to the already costly $(S/P)^2$ for planar images (where $P \geq 1$ denotes the patch length). This quickly becomes intractable when large resolutions are coupled with short patch lengths. Liu *et al.* [21] proposed the use of a *shifted-window* scheme to confine the attention computation to windows[†] while still allowing windows to exchange information amongst each other, reducing the sequence length to a mere $W^2$ (where $W \geq 1$ denotes the *window length* in patches). The appeal of this method is that the choice of $W$ is arbitrary and can thus be adjusted to make the model fit on the respective hardware. Naturally, this advantage becomes even more pronounced for volumetric data where the sequence length is given by $W^3$, a drastic improvement from $L = (S/P)^3$ as incurred by ViT. However, it is crucial to point out that this comes at the expense of a reduced context size, as the self-attention is computed over a *fixed* number of patches, making it necessary to interleave multiple layers with (two) different attention masks to increase the context size.

We propose a novel attention mechanism, G2L-MSA, to enable information exchange between windows that achieves a tight spatial packing of global information, re-envisioning the shifted-window attention mechanism. Fundamentally, this enables *unconstrained* window-to-window computation at each stage of the network, while featuring identical computational complexity as compared to both attention masks (*i.e.*W-MSA & SW-MSA) used by the Swin-Transformer. Additionally, G2L-MSA can be trivially extended to work on volumetric data such as MRI scans.

---

[†]A window is a grid of adjacent patches

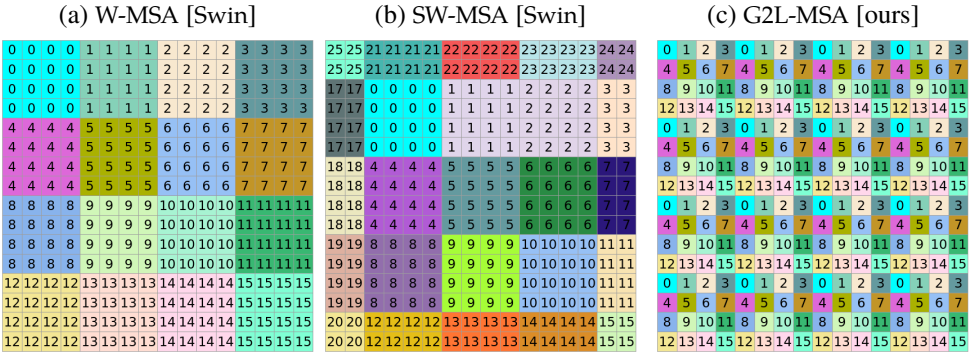

Figure 4: *Comparing attention masks of G2L-MSA to Swin (W-MSA and SW-MSA):* For the sake of simplicity we show the (2D) attention masks for a tensor of shape $M \times M$ for $M = {}^S/_P = 16$ and a window length of $W = 4$ patches (a) W-MSA performs self-attention over a contiguous window of $W \times W$ patches and does not lead to any window-to-window communication when iterated (b) Swin addresses this problem by letting newly defined windows cross the boundaries of former W-MSA windows (e.g. 0 connects windows $0, 1, 4$ and 5 as defined by W-MSA). Alternating between W-MSA and SW-MSA thus yields continually growing context sizes (c) G2L-MSA embraces non-contiguous windows and is a drop-in replacement to SW-MSA. Every newly defined window combines information from every previously defined W-MSA window, making the context size maximal after a single application while leaving the window size unchanged. The ViT architecture also achieves global communication yet at the expense of much longer and possibly intractable sequence lengths.

**G2L-MSA Definition** For a more clear illustration, we show how G2L-MSA is defined in a 2D-context, as it is straightforward to extend the construction to 3D. The window identifiers[‡] of W-MSA can be specified with the help of the *Kronecker product* as W-MSA$_{ID} = \mathcal{I} \otimes \mathbb{1}_{W \times W}$ where $\mathcal{I} = \text{Reshape}_{S/W \times S/W}(1, \ldots, (S/w)^2)$ and $\mathbb{1}_{W \times W}$ denotes a tensor of ones of shape $W \times W$. We define the window identifiers of G2L-MSA from those of W-MSA as follows:

$$\text{G2L-MSA}_{ID} = \text{G2L}_{W \times W}(\text{W-MSA}_{ID}) \tag{1}$$

where $\text{G2L}_{W_1 \times W_2} : \mathbb{R}^{S_1 \times S_2} \to \mathbb{R}^{S_1 \times S_2}$ denotes the *global-to-local* function (see Figure 3 for an example) for a block size of $W_1 \times W_2$ that we define as follows for some arbitrary matrix $Z \in \mathbb{R}^{S_1 \times S_2}$ and assuming that $W_1 \mid S_1$ and $W_2 \mid S_2$:

$$
\begin{bmatrix}
Z^{(1,1)} & \cdots & Z^{(1,S_2/W_2)} \\
\vdots & \ddots & \vdots \\
Z^{(S_1/W_1,1)} & \cdots & Z^{(S_1/W_1,S_2/W_2)}
\end{bmatrix}
\mapsto
\begin{bmatrix}
\begin{bmatrix}
z_{1,1}^{(1,1)} & \cdots & z_{1,1}^{(1,S_2/W_2)} \\
\vdots & \ddots & \vdots \\
z_{1,1}^{(S_1/W_1,1)} & \cdots & z_{1,1}^{(S_1/W_1,S_2/W_2)}
\end{bmatrix}
& \cdots &
\begin{bmatrix}
z_{1,W_2}^{(1,1)} & \cdots & z_{1,W_2}^{(1,S_2/W_2)} \\
\vdots & \ddots & \vdots \\
z_{1,W_2}^{(S_1/W_1,1)} & \cdots & z_{1,W_2}^{(S_1/W_1,S_2/W_2)}
\end{bmatrix} \\
\vdots & \ddots & \vdots \\
\begin{bmatrix}
z_{W_1,1}^{(1,1)} & \cdots & z_{W_1,1}^{(1,S_2/W_2)} \\
\vdots & \ddots & \vdots \\
z_{W_1,1}^{(S_1/W_1,1)} & \cdots & z_{W_1,1}^{(S_1/W_1,S_2/W_2)}
\end{bmatrix}
& \cdots &
\begin{bmatrix}
z_{W_1,W_2}^{(1,1)} & \cdots & z_{W_1,W_2}^{(1,S_2/W_2)} \\
\vdots & \ddots & \vdots \\
z_{W_1,W_2}^{(S_1/W_1,1)} & \cdots & z_{W_1,W_2}^{(S_1/W_1,S_2/W_2)}
\end{bmatrix}
\end{bmatrix}
\tag{2}
$$

where $\{Z^{(i,j)}\}_{i=1,\ldots,S_1/W_1, j=1,\ldots,S_2/W_2}$ denote the $W_1 \times W_2$ sized (non-overlapping) blocks of $Z$.

---

[‡] A window identifier tensor maps each element of the input feature map to a fixed set of windows. Elements with the same window identifiers are subject to the same self-attention computation.

```python
import numpy as np
def compute_g2l_msa_window_ids(S: int, W: int, D: int):
    num_windows_per_axis = S // W
    w_msa_window_ids = np.kron(a=np.reshape(np.arange(num_windows_per_axis ** D, dtype=np.int32),
                                            newshape=(num_windows_per_axis,) * D),
                               b=np.ones(shape=(W,) * D, dtype=np.int32))
    perm = [_ for pair in zip(range(1, 2 * D, 2), range(0, 2 * D, 2)) for _ in pair]  # transposes axes (2i, 2i+1)
    g2l_msa_window_ids = w_msa_window_ids\
        .reshape(*(num_windows_per_axis, W) * D)\
        .transpose(perm)\
        .reshape((S,) * D)
    return g2l_msa_window_ids
```

Figure 5: *Implementation* of G2L-MSA: The above code snippet creates the *D*-dimensional G2L-MSA$_{ID}$ window identifier tensor of shape $S \times \ldots \times S$ by transformation of the *D*-dimensional W-MSA window identifiers.

An implementation of G2L-MSA that works for an arbitrary number of dimensions can be found in Figure 5.

**G2L-MSA Efficiency properties** Because the function G2L$_{W_1 \times W_2}$ is defined as a spatial permutation of its input elements, we observe that the sequence length (= number of elements in a window) of the self-attention computations remains invariant for each window. This gives the same type of efficiency as W-MSA$_{ID}$ in Swin, so long as the attention calculations are performed on memory-adjacent patches. This type of memory "compactification" can be attained for G2L-MSA by first applying the *inverse* function of G2L$_{W \times W}$ (given by G2L$_{S/W \times S/W}$) on $X$, performing the self-attention computation, and finally applying G2L$_{W \times W}$ in order to move elements to their original locations.

**Architecture** Suppose that $X \in \mathbb{R}^{S \times S \times S}$ denotes an MR scan whose artifacts are to be filtered out, the architecture can be dissected into three components (i) For some patch length *P* dividing *S* the image is first partitioned into *volumetric* patches, yielding a tensor $X_P \in \mathbb{R}^{S/P, S/P, S/P, P^3}$ after flattening the patch dimensions. Each patch is subsequently linearly embedded to a feature size *E* using a shared learned projection $h_1$ resulting in $X_E \in \mathbb{R}^{S/P \times S/P \times S/P \times E}$ (ii) $X_E$ is passed through a residual chain (**"Residual Core Chain"**, see Figure 2) of interleaved (W-MSA) and G2L-MSA layers. We found normalization to be crucial to encourage convergence. Thus, *layer normalization* layers [1] are added before each attention block, before each head projection and before each fully-connected layer in the feature block§. (iii) Finally, the structural changes of (i) need to be reverted to obtain the image's original shape, accordingly a (shared) linear projection $h_2$ maps the vectors of size *E* back to the patch extent $P^3$ which ultimately enables reshaping to acquire $\tilde{X} \in \mathbb{R}^{S \times S \times S}$. An illustration of this process can be found in Figure 1. An algorithmic summary of the process is given as

$$X_P = \text{Patchify}_{3D}(X); \; X_E = h_1(\text{Flatten}_{3D}(X_P)); \; Z_0 = X_E$$

$$Z_l = Z_{l-1} + \begin{cases} g_{\theta_l}^{\text{W-MSA}}(Z_{l-1}), & \text{for odd } l \\ g_{\theta_l}^{\text{G2L-MSA}}(Z_{l-1}), & \text{for even } l \end{cases}; \; \tilde{X} = \text{Reshape}_{S \times S \times S}(\text{LayerNorm}(h_2(Z_L)))$$

where $l = 1, \ldots, L$ and $h_1 : \mathbb{R}^{P^3} \to \mathbb{R}^E$, $h_2 : \mathbb{R}^E \to \mathbb{R}^{P^3}$ and the individual layers $g_\bullet^{\text{W-MSA}}$ resp. $g_\bullet^{\text{G2L-MSA}}$ are realized as follows:

$$\mathcal{L} = \left\{ g_{\theta_l = \{\phi_l, \psi_l\}}^{\text{W-MSA}} = \mathcal{F}_{\phi_l} \circ \mathcal{A}_{\psi_l}^{\text{W-MSA}}; \; g_{\theta_l = \{\phi_l, \psi_l\}}^{\text{G2L-MSA}} = \mathcal{F}_{\phi_l} \circ \mathcal{A}_{\psi_l}^{\text{G2L-MSA}} \right\} \tag{3}$$

with $\mathcal{F}_\bullet$ denoting a feature projection sequence composed of *layer normalization – fully-connected ($E \mapsto 4E$) – GeLU [11] – layer normalization – fully-connected ($4E \mapsto E$)* layers.

---

§Feature blocks non-linearly map features from size *E* to size $4 \cdot E$ and finally back to *E*

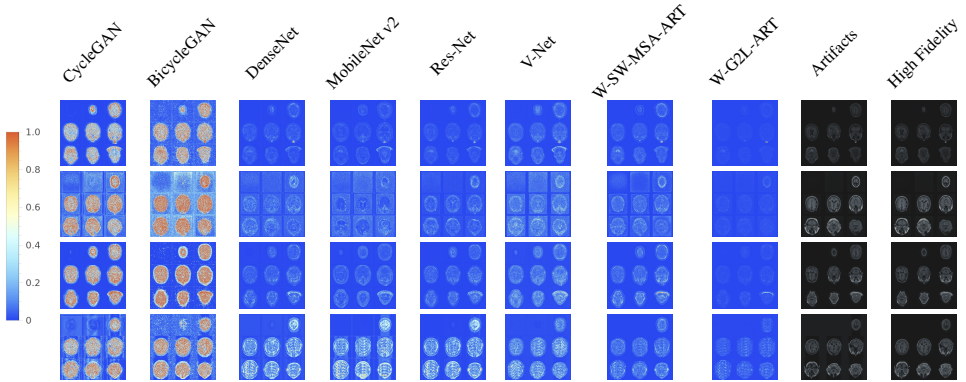

Figure 6: **Experiment I - Visual Quality** We compute *absolute difference* maps (for each column except *Artifacts* and *High Fidelity*) between reconstructions and their high-fidelity counterparts. Each row corresponds to one scan and values are normalized along rows to provide meaningful scaling. For the sake of illustration, we visualize each volume as a grid of $3 \times 3$ axial slices. The last column shows the high-fidelity original (no difference map).

Moreover, we make use of *relative* positional encoding within the self-attention module as suggested by Shaw *et al*. [25].

**Data & Artifact Simulation** Our supervised approach relies on two fundamental assumptions: First, the presence of high-fidelity data (or equiv. absence of artifacts), and secondly, the ability of the simulator to synthesize highly-realistic artifacts of sufficient diversity. We use the highly-curated and standardized ADNI [34] dataset to satisfy the first requirement. For the second, we utilize the TorchIO [23] MRI artifact ecosystem which integrates several state-of-the-art artifact simulators that synthesize *motion artifacts* [25], *random anisotropy* [2] or *bias field inhomogeneities* [26]. During training, we employ a *Bernoulli process* to sample artifacts in order to increase diversity. Suppose that $\mathcal{A} = (\mathcal{A}_1, \ldots, \mathcal{A}_K)$ denotes a sequence of artifact types and that $X$ denotes an artifact-free image. Artifact type $\mathcal{A}_1$ is applied with probability $p_1$ on $X$, artifact type $\mathcal{A}_2$ is then applied with prob. $p_2$ on the scan resulting from the last stage *irrespective* of its outcome. This process is iterated until all decisions are made. For the sake of simplicity[¶] we assume that $p_1 = \ldots = p_K = 1/K$.

## 4 Experiments

Through a series of experiments, we assess the quality of our reconstructions in comparison to numerous benchmark methods visually, according to standard summary metrics, and according to how well the artifact removal enables the succsss of critical downstream tasks.

**Benchmark Methods** We compare W-G2L-ART against multiple established neural network types. As our model is volumetric, we made necessary adaptations to these models (*e.g*. exchanging 2D convolutions with 3D convolutions) to enable fair comparisons, but otherwise kept changes minimal. To account for the small batch sizes necessitated by volumetric data, we choose to replace *batch normalization* [12] in these models with *instance normalization* [30] which is known to be unaffected by batch size. This results in more fair compar-isons between ViTs and CNNs. The methods we compare against include 3D versions of ResNet [9], DenseNet [11], V-Net [22], U-Net [24], and MobileNet-v2. CycleGAN [57] and BicycleGAN [58] are both *generative adversarial networks* [6] that aim to learn how to translate between two domains with the help of *cycle consistency* losses. As we have not been given access to the training code of DUNCAN, and given that this model operates on single slices of a different resolution, we refrain from incorporating DUNCAN into the comparision for the sake of fairness. As an ablation we also check how our model performs without G2L-MSA, *i.e.* by using a 3D variant of SW-MSA and call the resulting model W-SW-MSA-ART.

---

[¶]This implies that, in expectation, *one* artifact is present on each image as the number of applied artifacts is binomially distributed.

**Artifact Types** We use $K = 8$ different artifact types as defined by the TorchIO [23] library: *Random anisotropy* (downsampling followed by upsampling), *gamma expansion* (exponential change of contrast), *bias field inhomogeneities* (introduction of low-frequency intensity variations), *motion* (simulation of patient movement), *spiking* (spikes in $k$-space), *Gaussian blurring* and *additive Gaussian noise perturbation*. And finally, *ghosting*, which is due to semi-periodic movement. A visual depiction of the various artifact types can be found in the *Appendix*.

**Training** For all models we use the Adam optimizer [14] with a learning rate of $10^{-4}$ and $(\beta_1, \beta_2) = (0.9, 0.999)$. We use a batch size of 2 for resolution $128^3$ resp. 1 for resolution $256^3$. To reduce the variance of the gradients incurred by these small batch sizes we use 8-fold *gradient accumulation* to obtain effective batch sizes of 16 resp. 8. The models' complexities are maximized by adjusting the hyperparameters (*e.g.* number of layers/channels, kernel size) to fully utilize the memory of a NVIDIA 3090 RTX GPU. More details can be found in the *Appendix*. We use the ADNI dataset with a $80\% - 20\%$ training-test split.

**Experiment I - Qualitative assessment** We showcase the visual quality of the RAC models by sampling high-fidelity scans from the test set, applying the *artifact process* once on each of them, and then visualizing the *absolute differences* between the reconstructions and the high-fidelity scans. We observe that, subjectively, both W-SW-MSA-ART and W-G2L-ART seem to outperform the other methods, with W-G2L-ART having a minor edge over W-SW-MSA-ART. Four examples are provided in Figure 6, with additional examples provided in the Appendix.

**Experiment II - Quantitative assessment** We analyze how well the various models perform at the RAC task using standard measures for image reconstruction: PSNR and Structural Similarity (SSIM). Both metrics express how well the reconstruction resembles its high-fidelity counterpart. Results for PSNR can be found in Figure 7, while SSIM results are provided in Appendix Table 1 due to space constraints.

**Experiment III - Downstream task** To assess how well the various artifact removal approaches enable a downsteam task, we apply the skull-stripping software FSL BET [53] to corrected images and measure improvement according to DICE segmentation. Skull-stripping is a segmentation task that determines for each voxel whether it belongs to the skull or not. Suppose that $X_{\text{HF}}$ denotes a high-fidelity scan, $X_{\text{BAP}}$ the outcome of the *Bernoulli Artifact Process* when applied on $X_{\text{HF}}$ and $X_{\text{RAC}}$ the artifact-corrected version of $X_{\text{BAP}}$. We are interested in the *difference* of Dice scores defined as $\text{Dice}(X_{\text{HF}}, X_{\text{RAC}}) - \text{Dice}(X_{\text{HF}}, X_{\text{BAP}})$. Positive values indicate that applying RAC leads to improved skull-stripping segmentations and are thus preferred. These values are reported in Figure 7.

|  | PSNR ↑ | |
|---|---|---|
|  | $S = 128$ | $S = 256$ |
| BicycleGAN | $42.9 \pm 6.00$ | $45.6 \pm 6.00$ |
| CycleGAN | $43.7 \pm 6.75$ | $44.0 \pm 6.91$ |
| DenseNet | $52.6 \pm 12.52$ | $50.7 \pm 10.04$ |
| MobileNet | $54.2 \pm 12.33$ | $48.7 \pm 7.73$ |
| ResNet | $53.0 \pm 13.26$ | $51.5 \pm 10.40$ |
| V-Net | $51.1 \pm 11.58$ | $47.8 \pm 8.42$ |
| W-SW-MSA-ART | $51.1 \pm 12.22$ | $47.8 \pm 9.54$ |
| W-G2L-ART | $\mathbf{55.0 \pm 17.25}$ | $\mathbf{52.0 \pm 13.90}$ |

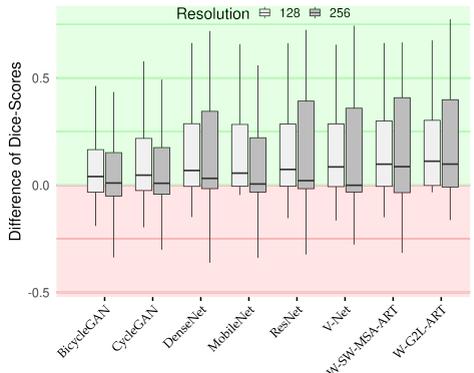

Figure 7: (left) **Exp. II - Qualitative assessment**: We simulate artifacts (on high-fidelity scans) using our *Bernoulli Artifact Process* on the test set and analyze how close (in terms of PSNR) the reconstructions of the models are in comparison to their high-fidelity counterparts. Higher values are preferable (right) **Exp. III - Downstream Task**: Positive values indicate that applying a RAC method on an artifact-affected scan increases the accuracy of the BET skull-stripping algorithm.

# 5   Discussion & Conclusion

In this work, we propose W-G2L-ART, a retrospective artifact correction (RAC) system using volumetric vision transformers with a novel global window-to-window communication mechanism. Our experiments show that W-G2L-ART is able to effectively correct eight different artifact types, compares favorably against the benchmark methods according to PSNR/SSIM, and gives the best results when corrected images are used for skull-stripping. Although there are conceptual advantages to the cycle-consistent learning approach used in CycleGAN and BicycleGAN, such as the freedom to use unpaired images, these approaches are susceptible to hallucinations that are discouraged in the more simple paired supervised setting we use. And since no large enough dataset with real artifacts exists, one is forced to synthesize artifacts, leaving little motivation to work with unpaired images – especially considering the additional complexity. Future work will address larger and more diverse datasets, including different modalities such as CT scans.

**Acknowledgements**: *This work was partially supported by the Swedish Research Council (VR) 2017-04609. Data collection and sharing for this project was funded by the Alzheimer's Disease Neuroimaging Initiative (ADNI) (National Institutes of Health Grant U01 AG024904) and DOD ADNI (Department of Defense award number W81XWH-12-2-0012). ADNI is funded by the National Institute on Aging, the National Institute of Biomedical Imaging and Bioengineering, and through generous contributions from the following: AbbVie, Alzheimer's Association; Alzheimer's Drug Discovery Foundation; Araclon Biotech; BioClinica, Inc.; Biogen; Bristol-Myers Squibb Company; CereSpir, Inc.; Cogstate; Eisai Inc.; Elan Pharmaceuticals, Inc.; Eli Lilly and Company; EuroImmun; F. Hoffmann-La Roche Ltd and its affiliated company Genentech, Inc.; Fujirebio; GE Healthcare; IXICO Ltd.; Janssen Alzheimer Immunotherapy Research & Development, LLC.; Johnson & Johnson Pharmaceutical Research & Development LLC.; Lumosity; Lundbeck; Merck & Co., Inc.; Meso Scale Diagnostics, LLC.; NeuroRx Research; Neurotrack Technologies; Novartis Pharmaceuticals Corporation; Pfizer Inc.; Piramal Imaging; Servier; Takeda Pharmaceutical Company; and Transition Therapeutics. The Canadian Institutes of Health Research is providing funds to support ADNI clinical sites in Canada. Private sector contributions are facilitated by the Foundation for the National Institutes of Health (www.fnih.org). The grantee organization is the Northern California Institute for Research and Education, and the study is coordinated by the Alzheimer's Therapeutic Research Institute at the University of Southern California. ADNI data are disseminated by the Laboratory for Neuro Imaging at the University of Southern California.*